\documentclass[letterpaper]{article}
\usepackage{aaai}
\usepackage{times}
\usepackage{helvet}
\usepackage{courier}
\usepackage{url}
\usepackage{graphicx}

\usepackage{caption}
\usepackage{subcaption}
 
\title{Confirmation detection in human-agent interaction using non-lexical speech cues}
\author{Mara Brandt, Britta Wrede, Franz Kummert and Lars Schillingmann\\ Cluster of Excellence Cognitive Interaction Technology (CITEC), Bielefeld University, 33615 Bielefeld, Germany \\ \texttt{\{mbrandt, bwrede, franz, lschilli\} at techfak.uni-bielefeld.de}}
 
\begin{document}
\maketitle

\begin{abstract}
Even if only the acoustic channel is considered, human communication is highly multi-modal. Non-lexical cues provide a variety of information such as emotion or agreement.
The ability to process such cues is highly relevant for spoken dialog systems, especially in assistance systems. In this paper, we focus on the recognition of non-lexical confirmations such as "mhm", as they enhance the system's ability to accurately interpret human intent in natural communication.
We implemented and evaluated a system for online detection of non-lexical confirmations. The architecture uses a Support Vector Machine to detect confirmations based on acoustic features. In a systematic comparison, several feature sets were evaluated for their performance on a corpus of human-agent interaction in a setting with naive users including elderly and cognitively impaired people. Our results show that using stacked formants as features yield an accuracy of 84\% outperforming regular formants and MFCC or pitch based features for online classification.
\end{abstract}

\section{Introduction}\label{sec:Introduction}

In human-machine interaction it is important to provide an intuitive interface for the users that allows them to make free use of the modality space. 
Among humans, speech is one of the most important modalities to communicate information, although non-verbal modalities such as gaze, gesture or action have been shown to be highly relevant for establishing common ground as well. 

Speech contains discourse particles or interjections which are important markers about the speaker's attitude \cite{Anderson2000}. 
These particles, that can be part of an utterance or also standalone (interjections), help to ground not only propositional meaning but also convey epistemic states \cite{Fischer2007}. 
Epistemic states pertain to the attitude of a speaker towards the information i.e. whether the speaker believes that the propositional content of an utterance (the own or the interlocutor's) is new and surprising or is already grounded, whether the speaker believes that the information is correct etc.
This information is highly relevant also in HCI which still tends to be quite brittle with respect to grounding. It is important to make use of these rather subtle cues to infer the user's attitude towards the current interaction. 
It is therefore important for a human-agent interaction to change the dialog structure according to the perceived internal state of the user by slowing down or repeating if uncertainty is perceived, or by continuing if the user is confirming.

However, discourse particles/markers or interjections have certain characteristics that render them difficult for automatic recognition with standard ASR approaches: 
For one, their use and surface structure is highly variable between different speakers \cite{Bell2001}.
Second, discourse particles are often characterized by stylized intonations, i.e. significantly different intonation patterns than ''normal'' speech \cite{Gibbon1997}.
Indeed, it has been shown that extreme values for prosodic features yield a much higher word error rate in ASR systems \cite{Jurafsky2010}
making it difficult to recognize the lexical units of discourse particles. But also, the meaning of discourse particles depends on the underlying intonation \cite{Gibbon1997}.
However, standard approaches to ASR do explicitly not take prosodic information into account.

In order to understand the meaning of the discourse particle, it is thus necessary to develop new approaches and investigate their acoustic nature.

%

%
%

One important feedback signal in dialogs is positive acknowledgment which indicates that the listener is still hearing and understanding what is being said. These feedback signals are often called ''filled-pauses'' and contain generally non-lexical acoustic units such as ''mhm'' or ''aha''. It has been shown that this feedback can be used by interaction partners to infer the listener's meta-cognitive state \cite{Brennan1995}.

While the phonetic realizations may be variable, it has been shown that their prosodic cues remain very stable with a very slowly and smoothly declining F0 \cite{Tsiaras}.
More specifically, fillers have a flat pitch, which lies in the median of pitch of the user across all his utterances \cite{Garg2006}.
Also, filled-pauses show a very specific articulation in that the articulators do not change their positions, yielding very stable formants and minimal coarticulation effects \cite{Audhkhasi2009}.
This is acoustically reflected in small fundamental frequency transitions and small spectral envelope deformations 
\cite{Goto1999}.

\section{Dataset}
\subsection{Scenario}
Our research is part of the KOMPASS project \cite{YaghoubzadehBuschmeier2015}. In this project, a virtual agent ``BILLIE'' is developed to help elderly and cognitively impaired people to plan and structure their daily activities, get reminders and suggestions for possible activities. The users interact naturally with the system to enter their appointments, therefore the system needs to understand natural language inputs and react to feedback. In addition to visual cues for understanding and confirmation, e.g. nodding, it is important for the dialog system to detect non-lexical confirmations like ``mhm'', because the automatic speech recognition (ASR) typically doesn't recognize them.
\subsection{User Study}
\begin{figure}[t]
	\centering
	\includegraphics[width=\columnwidth,height=\textheight,keepaspectratio]{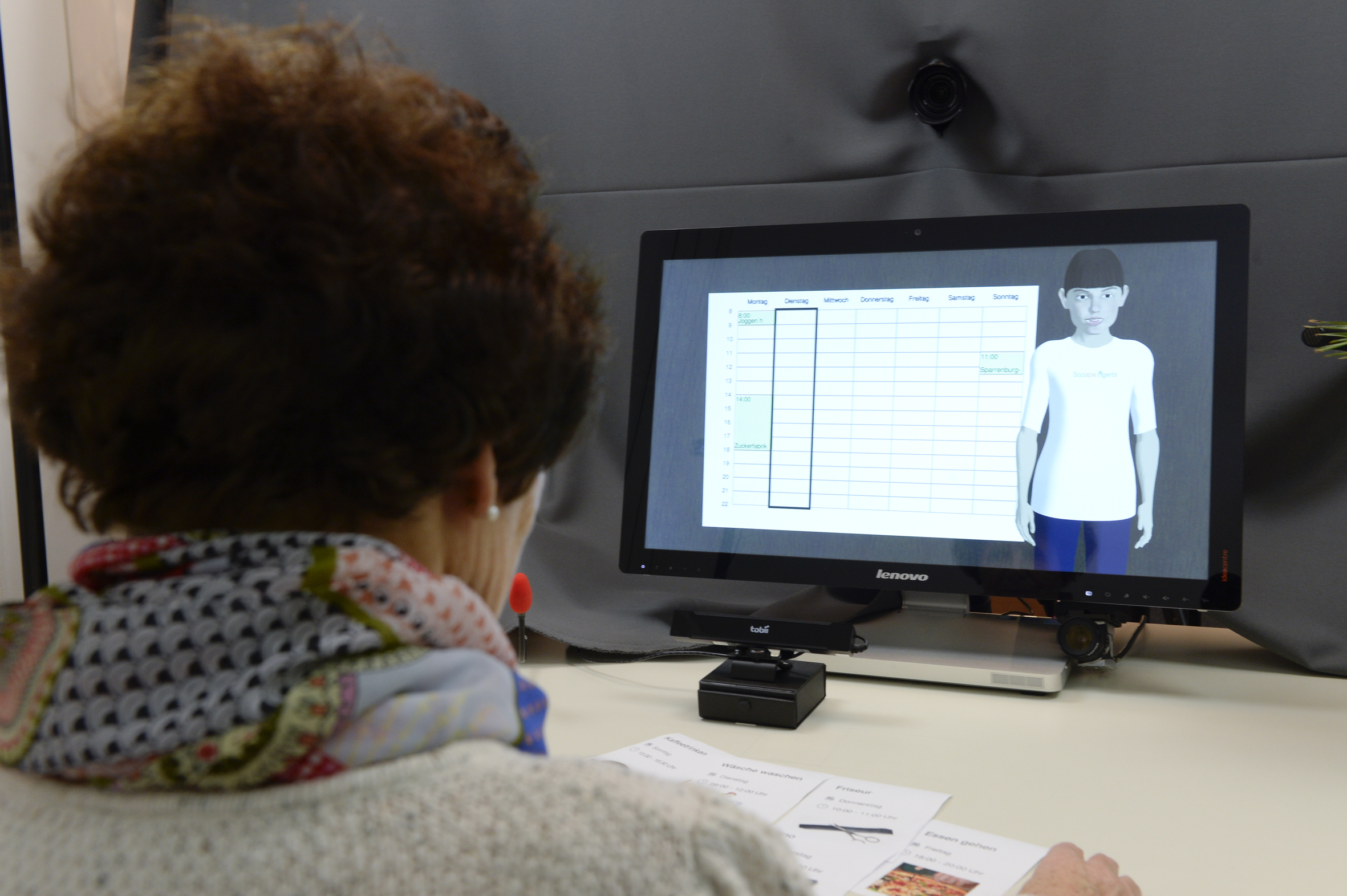}
	\caption{Interaction with the virtual agent ''BILLIE''}
	\label{fig:WOZ1-interaction}
\end{figure}

As part of the KOMPASS project a user study with participants of the intended user groups, elderly and cognitively impaired people, was conducted. The study was performed as a Wizard of Oz experiment \cite{Kelley1984}. 52 participants, consisting of 18 elderly (f: 14, m: 4), 18 cognitively impaired (f: 10, m: 8) and 16 students (f: 10, m: 6) with German as their first language, interacted with ``BILLIE`` and planned their daily activities for one week (Fig. \ref{fig:WOZ1-interaction}). The participants only got instructions to enter their appointments naturally in German without being instructed to use any special commands or phrases, resulting in natural communication with the system.

\subsection{Annotations}
The KOMPASS WOZ1 corpus was annotated automatically and manually. Voice activity detection (VAD) was used to divide the speech signal into segments of continuous speech. 
All segments are automatically annotated as regular utterances, unless they contain non-lexical confirmations, which were manually annotated. The distribution of regular utterances and non-lexical confirmations as well as the subsets used for this evaluation can be seen in Tab. \ref{tab:WOZ1-corpus}. 
The regular utterances contain 394 manually annotated filled-pauses, e.g. elongations and fillers, some of them similar to confirmations, e.g. ''hmm``, which can lead to false-positives in the detection of non-lexical confirmations. 
\begin{table}[t]
	\resizebox{\columnwidth}{!}{
	\begin{tabular}{*{5}{c}}
		\hline
		set & \#participants (f, m) & \#all segments & \#confirmations \\
		\hline
		WOZ1 data & 52 (f: 34, m: 18) & 5385 & 129 \\
		Training set & 17 (f: 14, m: 3) & 1885 & 87 \\
		Test set & 4 (f: 3, m: 1) & 415 & 42 \\
		\hline
	\end{tabular}
	}
	\caption{WOZ1 corpus segment distribution and used subsets for training with cross-validation and testing}
	\label{tab:WOZ1-corpus}
\end{table}
\section{Non-Lexical Confirmation Detection System}
\subsection{Architecture}
\begin{figure}[t]
	\centering
	\includegraphics[width=\columnwidth,height=\textheight,keepaspectratio]{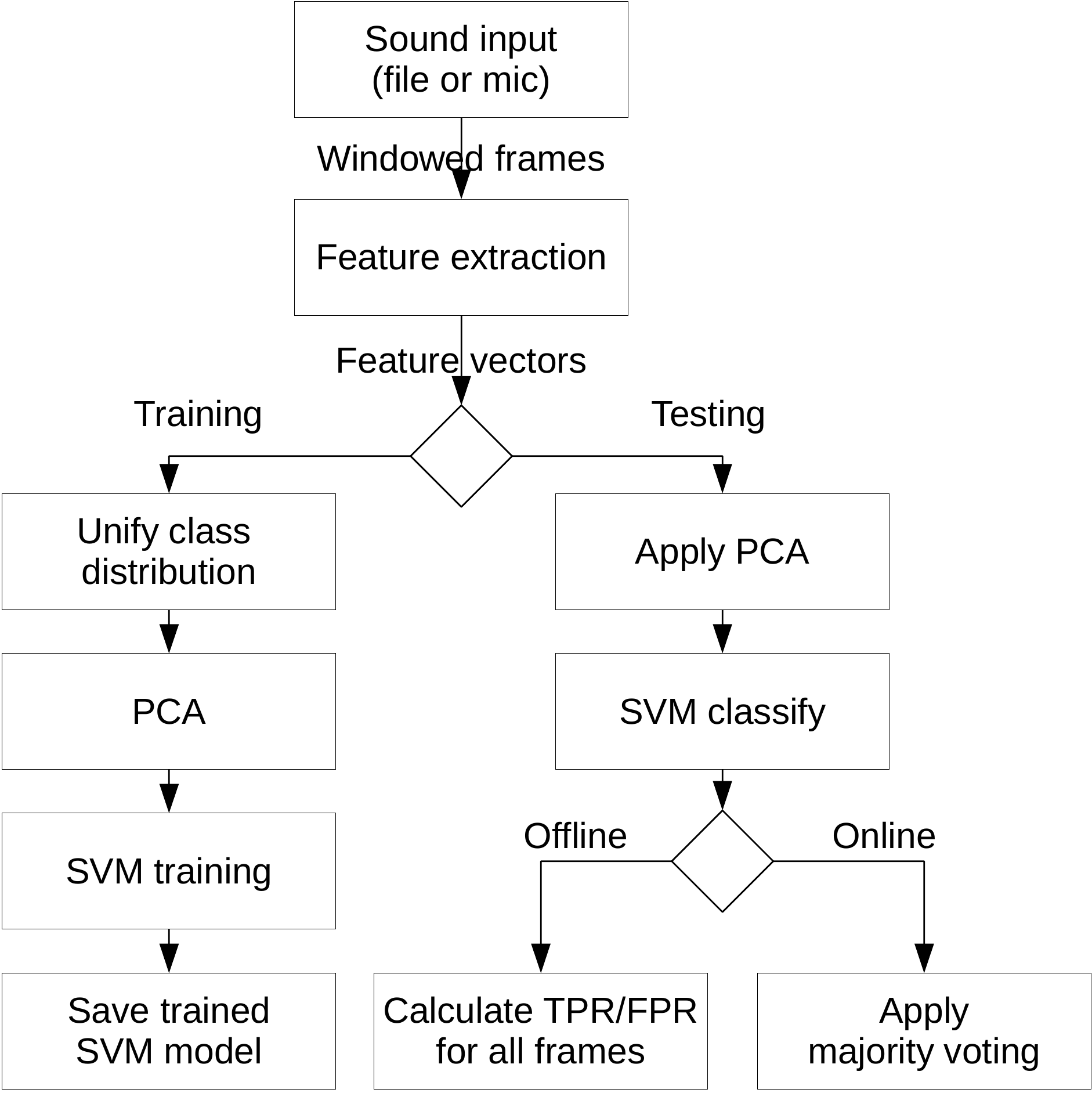}
	\caption{System architecture}
	\label{fig:system-diagram}
\end{figure}
\begin{figure}[t]
	\centering
	\includegraphics[width=\columnwidth,height=\textheight,keepaspectratio]{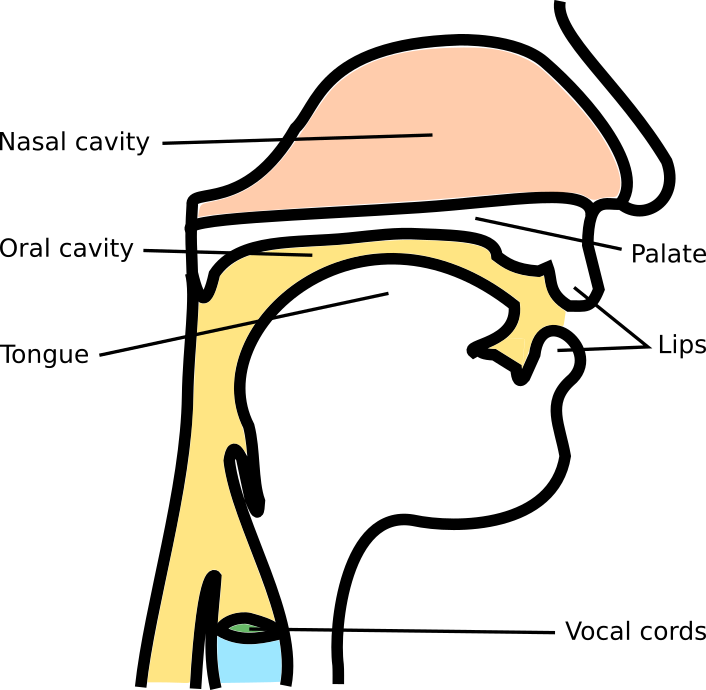}
	\caption{Source-filter model: vocal cords and vocal tract based on \cite{Philippsen}}
	\label{fig:Source-filter}
\end{figure}
We developed a system for the recognition of non-lexical confirmations that can handle sound input from files or microphone recording. The architecture of this system is shown in Fig. \ref{fig:system-diagram}.
To support both offline and online processing with mostly the same algorithms, the software consists of modules that can be used in both modes. 
The input, using either sound files or a microphone as source, is chunked into overlapping frames of 25ms with a frame shift of 10ms.
After that, the frames are windowed with a Blackman-Harris window \cite{Harris1978} for the MFCC related feature sets or a Hann window \cite{Blackman1959} for formant and pitch based feature sets and the different selected features are extracted frame by frame.
Principal Component Analysis (PCA) \cite{Pearson1901} is performed to reduce the dimensionality of the feature vectors of feature sets with stacked features or derivatives except for stacked formants, which is necessary because of the high dimensionality of the stacked features and the small amount of data.
In training mode, a Support Vector Machine (SVM) \cite{Vapnik1995} is trained as described in Sec. \ref{sec:SVM-training} and the trained model is serialized for later classification tasks. 
For the classification mode, the same steps are required, but instead of the SVM training, the sound input is classified frame by frame with the deserialized trained SVM model.
The classification results are calculated according to the description of offline and online classification in Sec. \ref{sec:classification}.
\begin{figure*}[t]
	\centering
	\begin{subfigure}{.5\textwidth}
	  \centering
	  \includegraphics[width=\columnwidth,height=\textheight,keepaspectratio]{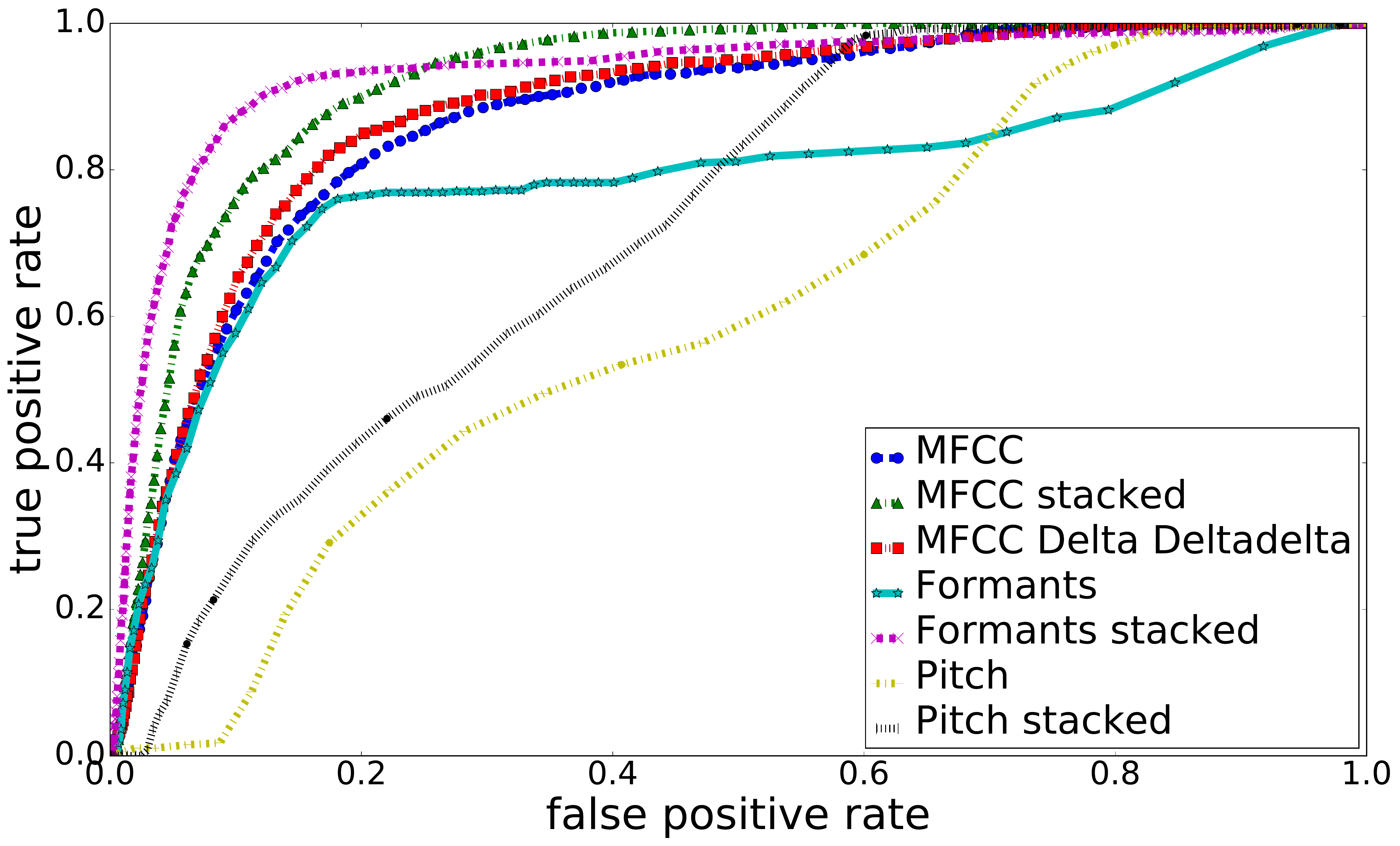}
	  \caption{ROC plot of the first test set}
	  \label{fig:sub1}
	\end{subfigure}%
	\begin{subfigure}{.5\textwidth}
	  \centering
	  \includegraphics[width=\columnwidth,height=\textheight,keepaspectratio]{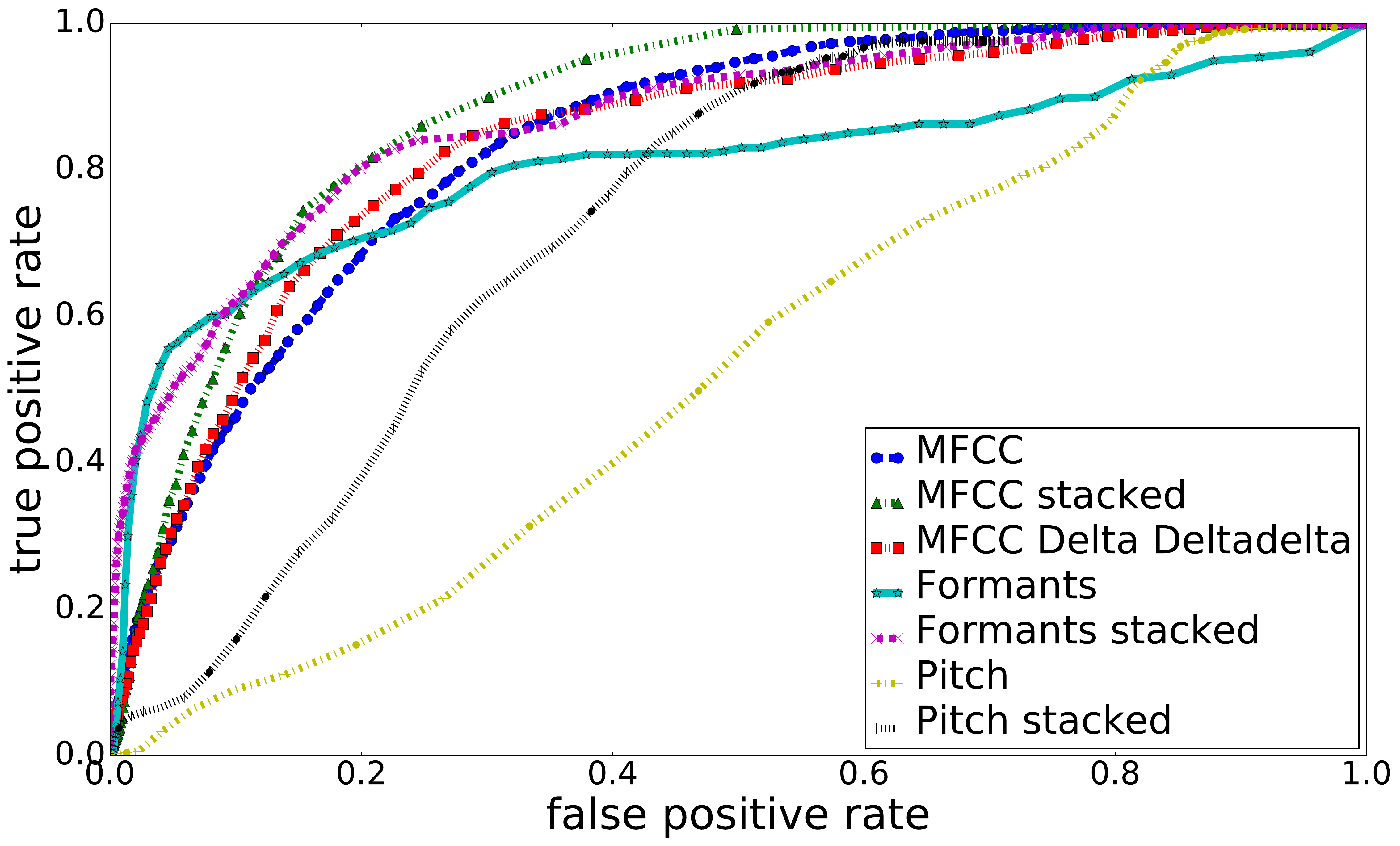}
	  \caption{ROC plot of the second test set}
	  \label{fig:sub2}
	\end{subfigure}
	\caption{ROC curves of classifiers with different feature sets}
	\label{fig:roc_curve}
\end{figure*}

\subsection{Feature Extraction/Selection}\label{sec:feature-extraction}
As argued above, different features may be suited for recognizing filled pauses. On the one hand, the pitch contour appears to be very salient, thus F0 becomes an interesting feature. On the other hand, the vocal tract (Fig. \ref{fig:Source-filter}) remains stable, which would be reflected in the formants and the MFCCs. The different features are described in this section and an overview of the resulting feature vectors with corresponding sizes are shown in Tab. \ref{tab:feature-vectors}.
\paragraph{Mel-Frequency Cepstral Coefficients}
Mel-frequency cepstral coefficients (MFCCs) are features common in speech recognition. They reflect properties of the vocal tract during speech production and mimic human perception of speech. The coefficients are designed to mitigate speaker dependent characteristics. MFCCs are extracted with the Essentia framework \cite{Bogdanov2013}. For this the sound signal is windowed with a Blackman-Harris window and the spectrum of this window is computed. After that, the first 13 MFCCs in the frequency range from 20 Hz to 7800 Hz are calculated with 40 mel-bands in the filter.
\paragraph{Differentiation}
To capture the salient articulation we also calculated the first and second derivatives of the MFCCs ($\Delta$ and $\Delta\Delta$). For this the polynomial filter introduced by Savitzky and Golay \cite{Savitzky1964} is used which combines differentiation and smoothing. The general formula for this filter is shown in Equ. \ref{equ:savitzky-golay-formular}, where $n$ is the filter length, $a_i$ are the coefficients and $h$ is the normalization factor.
\begin{equation}
y_t=\frac{1}{h}\sum_{i=-\frac{n-1}{2}}^{\frac{n-1}{2}}a_ix_{t+i}
\label{equ:savitzky-golay-formular}
\end{equation}
Savitzky and Golay provide coefficients to use for the calculation of the derivatives (Tab. \ref{tab:SG-coeffs}).

\begin{table}[h]
	\resizebox{\columnwidth}{!}{
	\begin{tabular}{*{4}{c}}
		\hline
		derivative & filter length & coefficients & h \\
		& & & (normalization factor) \\
		\hline
		first & 7 & -3, -2, -1, 0, 1, 2, 3 & 28\\
		\hline
		second & 7 & 5, 0, -3, -4, -3, 0, 5 & 42\\
		\hline
	\end{tabular}
	}
	\caption{Savitzky-Golay filter coefficients}
	\label{tab:SG-coeffs}
\end{table}
\paragraph{Stacked MFCCs}
Stacked MFCCs are another way to model the context and dynamics of MFCCs and can outperform MFCC derivatives \cite{Heck2013}. Instead of calculating the derivatives, the 13 MFCCs of adjacent frames are stacked to form a single feature vector. 15 stacked frames, resulting in a 195-dimensional feature vector, yield the best results in terms of true positive and false positive rate for non-lexical confirmation recognition.
\paragraph{Formants}
As formants are directly correlated with movements of the vocal tract, they should be able to provide good features for filled-pause detection (see Sec. \ref{sec:Introduction}). Non-lexical confirmations are very similar to some filled-pauses, so formants can be used for non-lexical confirmation detection. The linear predictive coding (LPC) algorithm of the Essentia framework is used to calculate linear predictive coefficients of the order 12. These coefficients are used to calculate the polynomial roots using the Eigen3 PolynomialSolver algorithm \cite{eigenweb}. Subsequently, the roots are fixed into the unit circle. The first two formant frequencies which show the configuration of the vocal tract are calculated from these fixed roots.
To measure the stability of the formants, the standard deviation of each formant over 15 frames is calculated and added as features.
\paragraph{Stacked Formants}
The idea of stacked MFCCs to model the dynamics of the signal over time can also be applied to formants. The 2 formants calculated per frame are therefore stacked over 15 frames to form one 30-dimensional feature vector.
\paragraph{Pitch}
Pitch is a feature that measures the frequency of the vocal cord vibrations.
It is calculated using the PitchYinFFT algorithm \cite{paul2007a} of the Essentia framework, which is an optimized version of the Yin algorithm calculated in the frequency domain. The input is therefore windowed with a Hann window.
\paragraph{Stacked Pitch}
Corresponding to the approach with MFCCs and Formants, the calculated pitch values over 15 frames are stacked to form one feature vector.
\paragraph{Principal Component Analysis}
Principal Component Analysis (PCA) is applied to the feature vectors of feature sets with stacked features or derivatives, except for stacked formants, to reduce dimensionality and to transform the features into linearly uncorrelated variables that describe the largest possible variances in the data. The algorithm used is the vector\_normalizer\_pca from the dlib library \cite{dlib09}. When the PCA is performed, the feature vectors are normalized automatically and no additional normalization is necessary.
The PCA parameter $\epsilon$ controls the size of the transformed feature vector. A value $0 < \epsilon <= 1$ can be chosen and large values result in larger feature vectors. For our evaluation, we chose a value of 0.95 to maintain most of the information contained in the features.
\begin{table}[h]
	\resizebox{\columnwidth}{!}{
	\begin{tabular}{ l c }
		\hline
		Feature combination & Feature vector size \\
		\hline
		MFCCs & 13 \\
		MFCCs Delta Deltadelta & 39 \\
		Stacked MFCCs (15 frames) & 195 \\
		SD of formants (15 frames) & 2 \\
		Stacked formants (15 frames) & 30 \\
		Pitch & 1 \\
		Stacked pitch (15 frames) & 15\\
		\hline
	\end{tabular}
	}
	\caption{Tested feature combinations with corresponding feature vector sizes (without PCA)}
	\label{tab:feature-vectors}
\end{table}
\subsection{Model estimation}\label{sec:SVM-training}
A Support Vector Machine (SVM) with radial basis function (RBF) kernel is used for the classification of non-lexical confirmations vs. other speech.
Therefore, the svm\_c\_trainer algorithm of the dlib library is used in the implementation.
For the SVM training a feature vector of the selected features for each frame of a segment in the training set is calculated.
All feature vectors are collected and the features are normalized as part of the PCA transformation.
The normalized feature vectors are then used to train the SVM model.
\subsection{Classification}\label{sec:classification}
Classification can be performed in offline and online mode.
In offline mode, the classification results for each frame are stored and finally combined to calculate accuracy and true/false positive rates.
For online classification, each frame is classified and the classification results are added as $1$ for confirmations and $-1$ for other utterances to a rolling mean over 5 frames, which is used to implement a simple majority vote. If a specified majority threshold is reached the current VAD segment is classified as an utterance containing a non-lexical confirmation.
\section{Evaluation}
\begin{table*}[t]
	\centering
	\begin{tabular}{l*{6}{c}}
		\hline
		feature set & feature vector size & cross-validation result & TPR (\%) & FPR (\%) & ROC AUC \\
		\hline
		MFCCs & 13 & 76.5 - 81.4 & 82.7 & 20.2 & 0.87\\
		MFCCs Delta Deltadelta & 22 & 80.0 - 85.7 & 85.7 & 21.4 & 0.88 \\
		Stacked MFCCs & 58 & \textbf{91.0 - 96.8} & 82.8 & 10.0 & \textbf{0.92} \\
		Formants & 2 & 73.8 - 78.5 & 77.1 & 27.8 & 0.78\\
		Stacked Formants & 30 & 83.3 - 88.1 & 91.9 & 16.8 & \textbf{0.93} \\
		Pitch & 1 & 11.7 - 17.6 & 99.8 & 92.0 & 0.60\\
		Stacked Pitch & 11 & 37.9 - 43.8 & 99.1 & 65.1 & 0.71 \\
		\hline
	\end{tabular}
	\caption{Evaluation results: Cross-validation shows the stable performance of stacked MFCCs, but stacked formants achieve the highest ROC AUC for the test set}
	\label{tab:eval-results}
\end{table*}
\subsection{KOMPASS WOZ1 Data Preparation}\label{sec:data-preparation}
The KOMPASS WOZ1 data is prepared for SVM training and offline classification by extracting all VAD segments from the sound files. 
If the interval contains a manual annotation of a non-lexical confirmation, the segment is added as belonging to the non-lexical confirmation class. 
A script is used to split all segments in training and test sets. Users without non-lexical confirmation utterances are excluded.
We made sure to assign all utterances from one speaker to one of the sets only in order to achieve a speaker independent test set-up.
70 percent of the KOMPASS WOZ1 data are used for training, while the remaining 30 percent are used as test data (see Tab. \ref{tab:WOZ1-corpus}).
A leave-one-user-out cross-validation is performed on the training set. For each fold, all utterances of one user are used as the test set, while the remaining users are used for training. 
Because segments of non-lexical confirmations are usually shorter than other utterances, an additional step to maintain a uniform distribution of frames belonging to each class has to be performed. Feature vectors of other utterances are discarded prior to SVM training to compensate for the small number of frames belonging to non-lexical confirmations.
The test set contains a realistic subset of unevenly distributed utterances of both classes and all frames of these utterances are classified without balancing the uneven distribution of feature vectors of both classes.
\subsection{Parameter Optimization}\label{sec:parameter-optimization}
\begin{table}[h]
	\resizebox{\columnwidth}{!}{
	\begin{tabular}{l*{3}{c}}
		\hline
		Feature combination & $C$ & $\epsilon$ & $\gamma$ \\
		\hline
		MFCCs & 1 & 0.5 & 0.005 \\
		MFCCs Delta Deltadelta & 1 & 0.1 & 0.005 \\
		Stacked MFCCs (15 frames) & 1 & 0.5 & 0.005 \\
		SD of formants (15 frames) & 5 & 0.005 & 0.05 \\
		Stacked formants (15 frames) & 1 & 0.5 & 0.05 \\
		Pitch & 5 & 0.005 & 0.05 \\
		Stacked pitch (15 frames) & 5 & 0.5 & 0.05\\
		\hline
	\end{tabular}
	}
	\caption{Best SVM parameters found with grid search for each feature set}
	\label{tab:grid-search-results}
\end{table}
Grid search was used to optimize the SVM parameters $C$, $\epsilon$ and $\gamma$ for the RBF kernel. The parameters were tested in the ranges $C\in\{1, 5\}$, $\epsilon\in\{0.005, 0.05, 0.1, 0.5\}$ and $\gamma\in\{0.005, 0.05\}$. The best results for each feature set are shown in Tab. \ref{tab:grid-search-results}.
\subsection{Results on the KOMPASS WOZ1 Data}
The system for non-lexical confirmation detection was tested on the KOMPASS WOZ1 data. Seven different feature sets were evaluated: MFCCs, MFCCs + first and second derivative ($\Delta$, $\Delta\Delta$), stacked MFCCs, formants, stacked formants, pitch and stacked pitch. Grid search was performed as described in Sec. \ref{sec:parameter-optimization} for parameter optimization.
Before the SVM was trained, a leave-one-user-out cross-validation was performed (see Sec. \ref{sec:data-preparation}). To evaluate the performance of the trained models, the sum of the accuracy value weighted with the number of non-lexical confirmations for each fold was calculated. 

Fig. \ref{fig:roc_curve} shows the Receiver Operating Characteristic (ROC) curves of the seven classifiers with different feature sets, that were evaluated on different test sets. For the first test set, the stacked formants can outperform all other feature sets with an area under the curve (AUC) of 0.93. 
In comparison, the standard deviation of the formants achieve an AUC of 0.78, which is even below all of the MFCC related feature sets. 
The feature vectors consisting of 13 MFCCs result in classification results with an AUC of 0.87 
and adding first and second derivative only slightly improves the result (AUC of 0.88).
Stacking of the MFCCs raises the AUC to 0.92
, but stays below the value of stacked formants.
Using pitch as a single feature results in a nearly diagonal ROC curve (AUC of 0.60
), which corresponds to classification results near chance level. Stacking the pitch values to feature vectors over 15 frames only slightly improves the results (AUC of 0.71
).
The results with the second test set show, that the performance of the formant related feature sets is not stable, while the results of MFCC related and pitch related feature sets remain similar.

The online classification was evaluated with the two best feature sets, stacked formants and stacked MFCCs, which achieve accuracy values of 84\% and 79\%, respectively.
\section{Discussion}
In this paper, we described a system for non-lexical confirmation detection in speech. 
Our system is capable of both online and offline processing of speech data. Thus, it can easily be integrated into systems interacting with humans. 
We relied on Support Vector Machines with a RBF kernel for classification. A sliding window approach enables the system to spot filled-pauses within utterances without the necessity to explicitly model parts of speech not relevant for filled-pause detection. 
The system's performance was evaluated on seven different feature sets: MFCCs, MFCCs with first and second derivative ($\Delta$, $\Delta\Delta$), stacked MFCCs, formants, stacked formants, pitch and stacked pitch. 
The results show that successfully detecting non-lexical confirmations requires several frames of context. For this the stacking of the features improves the results and outperforms feature sets with derivatives. 
The results with stacked MFCCs, and MFCC related feature sets in general, are more stable within several performed test runs. But stacked formants have the potential to achieve higher classification results depending on the data. The amount of available data for SVM training might also influence the performance of the stacked feature sets and has to be evaluated.

Our approach can be applied to spot other acoustic events in speech data. In further studies, we aim to apply stacked features for the detection of other non-lexical speech events such as filled-pauses and for detecting socio-emotional signals such as uncertainty. Virtual agents like ''BILLIE`` will become more and more natural interaction partners by integrating those cues.
\section*{Acknowledgments}
The authors gratefully acknowledge the German Federal Ministry of Education and Research (BMBF) for providing funding to project KOMPASS (FKZ 16SV7271K),
within the framework of which our research was able to take place. 
This work was supported by the Cluster of Excellence Cognitive Interaction Technology ``CITEC'' (EXC 277) at Bielefeld University, which is funded by the German Research Foundation (DFG). 
Furthermore, the authors would like to thank our student worker Kirsten K\"astel for data annotation.
\bibliographystyle{aaai}
\bibliography{library}

\begin{thebibliography}{}

\bibitem[\protect\citeauthoryear{Andersen and Thorstein}{2000}]{Anderson2000}
Andersen, G., and Thorstein, F.
\newblock 2000.
\newblock {\em Pragmatic Markers and Propositional Attitude}.
\newblock Amsterdam: John Benjamins.
\newblock chapter Introduction,  1--16.

\bibitem[\protect\citeauthoryear{Audhkhasi \bgroup et al\mbox.\egroup
  }{2009}]{Audhkhasi2009}
Audhkhasi, K.; Kandhway, K.; Deshmukh, O.~D.; and Verma, A.
\newblock 2009.
\newblock Formant-based technique for automatic filled-pause detection in
  spontaneous spoken english.
\newblock In {\em 2009 IEEE International Conference on Acoustics, Speech and
  Signal Processing},  4857--4860.

\bibitem[\protect\citeauthoryear{Bell, Boye, and Gustafson}{2001}]{Bell2001}
Bell, L.; Boye, J.; and Gustafson, J.
\newblock 2001.
\newblock Real-time handling of fragmented utterances.
\newblock In {\em Proc. NAACL workshop on adaptation in dialogue systems},
  2--8.

\bibitem[\protect\citeauthoryear{Blackman and Tukey}{1959}]{Blackman1959}
Blackman, R.~B., and Tukey, J.~W.
\newblock 1959.
\newblock {\em Particular Pairs of Windows}.
\newblock Dover.
\newblock  98--99.

\bibitem[\protect\citeauthoryear{Bogdanov \bgroup et al\mbox.\egroup
  }{2013}]{Bogdanov2013}
Bogdanov, D.; Wack, N.; G\'{o}mez, E.; Gulati, S.; Herrera, P.; Mayor, O.;
  Roma, G.; Salamon, J.; Zapata, J.; and Serra, X.
\newblock 2013.
\newblock Essentia: An open-source library for sound and music analysis.
\newblock In {\em Proceedings of the 21st ACM International Conference on
  Multimedia}, MM '13,  855--858.
\newblock New York, NY, USA: ACM.

\bibitem[\protect\citeauthoryear{Brennan and Williams}{1995}]{Brennan1995}
Brennan, S.~E., and Williams, M.
\newblock 1995.
\newblock {The feeling of another's knowing: Prosody and filled pauses as cues
  to listeners about the metacognitive states of speakers}.

\bibitem[\protect\citeauthoryear{Brossier}{2007}]{paul2007a}
Brossier, P.~M.
\newblock 2007.
\newblock {\em Automatic Annotation of Musical Audio for Interactive
  Applications}.
\newblock Ph.D. Dissertation, Queen Mary, University of London.

\bibitem[\protect\citeauthoryear{Fetzer and Fischer}{2007}]{Fischer2007}
Fetzer, A., and Fischer, K.
\newblock 2007.
\newblock {\em Lexical Markers of Common Grounds}.
\newblock London: Elsevier.
\newblock chapter Introduction,  12--13.

\bibitem[\protect\citeauthoryear{Garg and Ward}{2006}]{Garg2006}
Garg, G., and Ward, N.
\newblock 2006.
\newblock {Detecting Filled Pauses in Tutorial Dialogs}.
\newblock (0415150):1--9.

\bibitem[\protect\citeauthoryear{Gibbon and Sassen}{1997}]{Gibbon1997}
Gibbon, D., and Sassen, C.
\newblock 1997.
\newblock Prosody-particle pairs as dialogue control signs.
\newblock In {\em Proc. Eurospeech}.

\bibitem[\protect\citeauthoryear{Goldwater, Jurafsky, and
  Manning}{2010}]{Jurafsky2010}
Goldwater, S.; Jurafsky, D.; and Manning, C.~D.
\newblock 2010.
\newblock Which words are hard to recognize? prosodic, lexical, and disfluency
  factors that increase speech recognition error rates.
\newblock {\em Speech Communication} 52(3):181 -- 200.

\bibitem[\protect\citeauthoryear{Goto, Itou, and Hayamizu}{1999}]{Goto1999}
Goto, M.; Itou, K.; and Hayamizu, S.
\newblock 1999.
\newblock {A real-time filled pause detection system for spontaneous speech
  recognition}.
\newblock {\em Proceedings of EUROSPEECH}  227--230.

\bibitem[\protect\citeauthoryear{Guennebaud, Jacob, and
  others}{2010}]{eigenweb}
Guennebaud, G.; Jacob, B.; et~al.
\newblock 2010.
\newblock Eigen v3.
\newblock http://eigen.tuxfamily.org.

\bibitem[\protect\citeauthoryear{Harris}{1978}]{Harris1978}
Harris, F.~J.
\newblock 1978.
\newblock On the use of windows for harmonic analysis with the discrete fourier
  transform.
\newblock {\em Proceedings of the IEEE} 66:51--83.

\bibitem[\protect\citeauthoryear{Heck \bgroup et al\mbox.\egroup
  }{2013}]{Heck2013}
Heck, M.; Mohr, C.; St{\"u}ker, S.; M{\"u}ller, M.; Kilgour, K.; Gehring, J.;
  Nguyen, Q.~B.; Nguyen, V.~H.; and Waibel, A.
\newblock 2013.
\newblock {\em Segmentation of Telephone Speech Based on Speech and Non-speech
  Models}.
\newblock Cham: Springer International Publishing.
\newblock  286--293.

\bibitem[\protect\citeauthoryear{Kelley}{1984}]{Kelley1984}
Kelley, J.~F.
\newblock 1984.
\newblock An iterative design methodology for user-friendly natural language
  office information applications.
\newblock {\em ACM Trans. Inf. Syst.} 2(1):26--41.

\bibitem[\protect\citeauthoryear{King}{2009}]{dlib09}
King, D.~E.
\newblock 2009.
\newblock Dlib-ml: A machine learning toolkit.
\newblock {\em Journal of Machine Learning Research} 10:1755--1758.

\bibitem[\protect\citeauthoryear{Pearson~FRS}{1901}]{Pearson1901}
Pearson~FRS, K.
\newblock 1901.
\newblock Liii. on lines and planes of closest fit to systems of points in
  space.
\newblock {\em Philosophical Magazine} 2(11):559--572.

\bibitem[\protect\citeauthoryear{Philippsen and Wrede}{2017}]{Philippsen}
Philippsen, A., and Wrede, B.
\newblock 2017.
\newblock {Towards Multimodal Perception and Semantic Understanding in a
  Developmental Model of Speech Acquisition}.
\newblock Presented at the 2nd Workshop on Language Learning at Intern. Conf.
  on Development and Learning (ICDL-Epirob) 2017.

\bibitem[\protect\citeauthoryear{Savitzky and Golay}{1964}]{Savitzky1964}
Savitzky, A., and Golay, M. J.~E.
\newblock 1964.
\newblock Smoothing and differentiation of data by simplified least squares
  procedures.
\newblock {\em Analytical Chemistry} 36(8):1627--1639.

\bibitem[\protect\citeauthoryear{Tsiaras, Panagiotakis, and
  Stylianou}{2009}]{Tsiaras}
Tsiaras, V.; Panagiotakis, C.; and Stylianou, Y.
\newblock 2009.
\newblock Video and audio based detection of filled hesitation pauses in
  classroom lectures.
\newblock In {\em 2009 17th European Signal Processing Conference},  834--838.

\bibitem[\protect\citeauthoryear{Vapnik}{1995}]{Vapnik1995}
Vapnik, V.~N.
\newblock 1995.
\newblock {\em The Nature of Statistical Learning Theory}.
\newblock New York, NY, USA: Springer-Verlag New York, Inc.

\bibitem[\protect\citeauthoryear{Yaghoubzadeh, Buschmeier, and
  Kopp}{2015}]{YaghoubzadehBuschmeier2015}
Yaghoubzadeh, R.; Buschmeier, H.; and Kopp, S.
\newblock 2015.
\newblock Socially cooperative behavior for artificial companions for elderly
  and cognitively impaired people.
\newblock In {\em Proceedings of the 1st International Symposium on
  Companion-Technology},  15--19.

\end{thebibliography}
\end{document}